\documentclass[10pt,doublecolumn,twoside]{IEEEtran}
\IEEEoverridecommandlockouts
\usepackage{cite}
\usepackage{amsmath,amssymb,amsfonts}
\usepackage{algorithmic}
\usepackage{graphicx}
\usepackage{textcomp}
\usepackage{epstopdf}
\usepackage{float}
\usepackage{esint}
\usepackage{subfigure}
\usepackage{color}
\usepackage{mathrsfs}
\usepackage{changes}
\usepackage{threeparttable}
\usepackage{makecell}

\def\BibTeX{{\rm B\kern-.05em{\sc i\kern-.025em b}\kern-.08em
    T\kern-.1667em\lower.7ex\hbox{E}\kern-.125emX}}

\makeatletter

\newcommand{\Rmnum}[1]{\expandafter\@slowromancap\romannumeral #1@}
\makeatother

\newcommand{\ls}[1]
    {\dimen0=\fontdimen6\the\font
     \lineskip=#1\dimen0
     \advance\lineskip.5\fontdimen5\the\font
     \advance\lineskip-\dimen0
     \lineskiplimit=.9\lineskip
     \baselineskip=\lineskip
     \advance\baselineskip\dimen0
     \normallineskip\lineskip
     \normallineskiplimit\lineskiplimit
     \normalbaselineskip\baselineskip
     \ignorespaces
    }

\begin{document}
\title{Redefining Information Freshness: AoGI for Generative AI in 6G Networks}

\author{Yuquan Xiao,~\IEEEmembership{Graduate Student Member,~IEEE}, Qinghe Du,~\IEEEmembership{Member,~IEEE}, Wenchi Cheng,~\IEEEmembership{Senior Member,~IEEE}, George K. Karagiannidis,~\IEEEmembership{Fellow,~IEEE}, Arumugam Nallanathan,~\IEEEmembership{Fellow,~IEEE}, and Mohsen Guizani,~\IEEEmembership{Fellow,~IEEE}
\thanks{Yuquan Xiao and Qinghe Du are with the School of Information and Communications Engineering, Xi'an Jiaotong University, Xi'an 710049, China. (e-mail:yqxiao@stu.xjtu.edu.cn,duqinghe@mail.xjtu.edu.cn)

Wenchi Cheng is with the State Key Laboratory of Integrated Services Networks, Xidian University, Xi'an 710071, China.

G. K. Karagiannidis is with Department of Electrical and Computer Engineering, Aristotle University of Thessaloniki, Greece.

Arumugam Nallanathan is with the School of Electronic Engineering and Computer Science, Queen Mary University of London, London E1 4NS, U.K.

M. Guizani is with the Machine Learning Department, Mohamed Bin Zayed University of Artificial Intelligence (MBZUAI), Abu Dhabi, UAE.}
\vspace{-20pt}
}

\maketitle

\begin{abstract}
Generative Artificial Intelligence (GenAI) is playing an increasingly important role in enriching and facilitating human life by generating various useful information, of which real-time GenAI is a significant part and has great potential in applications such as real-time robot control, automated driving, augmented reality, etc. There are a variety of information updating processes in real-time GenAI, and the age of information (AoI) is an effective metric for evaluating information freshness. However, due to the diversity and generativity of information in real-time GenAI, it may be incompatible to directly use existing information aging metrics to assess its timeliness. In this article, we  introduce a new concept called Age of Generative Information (AoGI) to evaluate the freshness of generative information, which takes into account the information delay caused not only by sampling and transmission, but also by computation. Furthermore, since real-time GenAI services are often supported by mobile-edge-cloud (MEC) collaborative computing in 6G networks and some of the generated information is privacy sensitive, it is recommended that the identities of edge and cloud should always be verified in a zero-trust manner. We introduce the concept of Age of Trust (AoT) to characterise the decay process of their trust level. We also discuss the optimisations of these evolved information aging metrics, focusing on the impact of dynamic external conditions, including wireless environments and limited computational resources. Finally, we highlight several open challenges in providing timeliness guarantees for real-time GenAI services.
\end{abstract}


\section{Introduction}
Generative Artificial Intelligence (GenAI) serves as a game-changing enabler to perform various complex tasks like humans, and is even expected to imbue humanoid robots with a soul, having attracted considerable attention recently~\cite{Xu2024Unleashing,Qu2025Mobile}. In particular, real-time GenAI, as a subset of GenAI, has broad application prospects in numerous interactive scenarios, such as augmented reality and automated driving. How to efficiently enable real-time GenAI services becomes a critical issue.

\begin{figure*}
  \centering
  \includegraphics[scale=0.67]{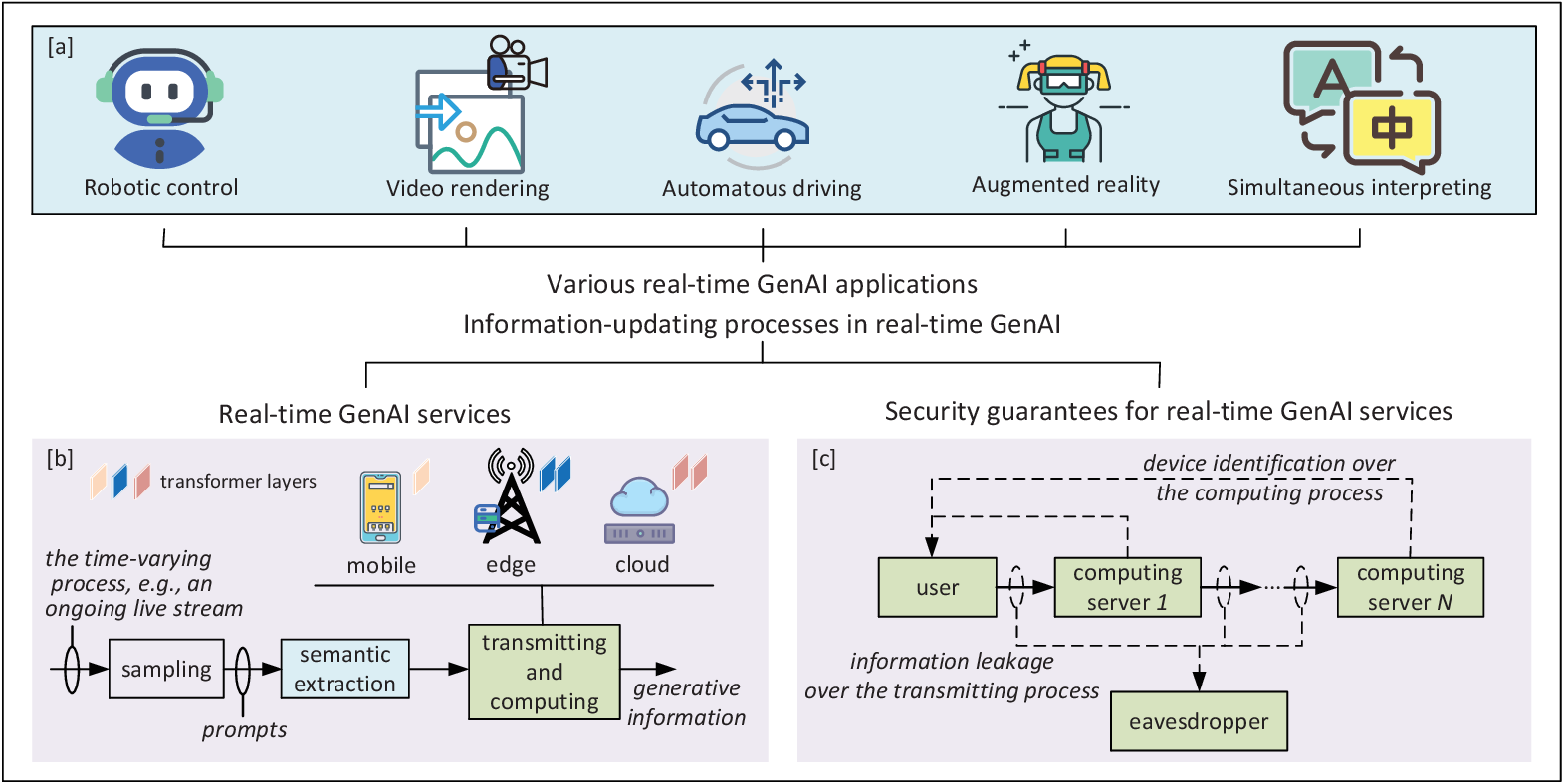}
  \caption{The information-updating processes in real-time GenAI services.}\label{fig:system_diagram}
\end{figure*}
Today, GenAI services are implemented using transformer-based or diffusion models with billions of neural parameters, where the inference and training processes are computationally intensive. Therefore, various GenAI models opt to be deployed on high-performance cloud servers for efficient computing, and then local devices request the GenAI services in a mobile-cloud manner. However, this setting may correspond to a long response delay, especially when network congestion occurs, and thus cannot meet the stringent requirements of various real-time GenAI applications~\cite{Zhang2024Mobile}. Compared with the mobile-cloud pattern, mobile-edge-cloud (MEC) collaborative computing is becoming a competitive stander for real-time GenAI services, where edge servers play an important role in relieving the pressure of data communication and task computation to cloud servers, and then it is not necessary to always upload raw data to cloud servers, thereby reducing the response delay. Accordingly, compared with the mobile-cloud pattern, the risk of information leakage when using MEC collaborative computing has the potential to be reduced. This is because data may only need to be exchanged between mobile devices and edge servers.

Although several efforts have been made to facilitate real-time GenAI services via MEC collaborative computing~\cite{Zhang2025Beyond}, most of them focus on end-to-end (E2E) delay minimization, which is suitable for query-based applications. Different from these, there is a category of real-time GenAI applications, such as view generation in augmented reality, which focus on a continuous information generation process, i.e., like an information updating process, and thus the existing delay-oriented solutions might no longer be applicable. Instead, it is more appropriate to use the metrics in the information updating domain, that is, the age of information (AoI)~\cite{Kaul2012Real}. AoI can capture the effects of sampling rate and transmission delay on the freshness of information, and has been demonstrated to be an effective metric for various real-time information updating systems. Notably, in real-time GenAI applications, information is not only sampled and transmitted, but also computed. Motivated by this observation, in this article we aim to build an information freshness evaluation framework for real-time GenAI. Our main contributions are summarized below:
\begin{itemize}
  \item We investigate the evolution of information aging metrics towards real-time GenAI. First, we present the typical information generation and updating processes in real-time GenAI. We then introduce several new metrics, including Age of Generative Information (AoGI) and Age of Trust (AoT), that fully account for the characteristics of these processes in real-time GenAI.
  \item The optimizations of the developed information aging metrics are discussed with an emphasis on AoGI, where it is elaborated how to minimize AoGI from a long-term perspective when facilitating real-time GenAI services via MEC collaborative computing. In particular, for risk-sensitive GenAI applications, we present a new concept called statistical AoGI, which allows fine-tuning the failure probability of peak AoGI to meet various requirements of risk-sensitive GenAI applications.
  \item The advantages of the AoGI-oriented solution compared to the delay-oriented solution are shown with a demonstration case where real-time GenAI services are supported by mobile edge collaborative computing. The simulation result shows that the AoGI-oriented solution can further improve the freshness of generative information by 25\%.
\end{itemize}

The rest of this article is organized as follows. Section~\ref{sec:info_process} presents the typical information-updating processes in real-time GenAI. Section~\ref{sec:evolution_aoi} investigates the evolution of information-aging metrics towards the above-mentioned processes. The optimizations on the evolved information-aging metrics are discussed in Section~\ref{sec:opt_aoi}. Finally, the article is concluded with open challenges in Section~\ref{sec:conclusion}.

\section{Typical Information-Updating Processes in Real-Time GenAI}
\label{sec:info_process}
Real-time GenAI involves several information updating processes that differ significantly from those in traditional sensor networks. While the information updates in sensor networks mainly focus on minimizing the data communication overhead to ensure its freshness, the computational overhead, such as inference and training costs, becomes the dominant factor in real-time GenAI. Furthermore, to avoid redundant computations, semantic extraction before inference is essential in real-time GenAI, which results in a more flexible information update process. In addition to maintaining the freshness of generative information in real-time GenAI, there are also security concerns, such as preventing the leakage of real-time generative information and verifying the identity of service providers. These challenges require a fundamental design to guide the deployment of real-time GenAI services.

\subsection{Real-Time GenAI Services}
As shown in Fig.~\ref{fig:system_diagram}(a), real-time GenAI is envisioned to enable various potential applications. For example, real-time robot control using GenAI is expected to timely sense the environment and comprehensively understand human intention, and then make appropriate human-like responses~\cite{lin2023pushing}, which has significant potential in the field of service industries. Real-time GenAI can also be applied to video rendering and its style transfer, which is beneficial to live streaming and online shopping. In addition, real-time GenAI has been considered as an enabler for intelligent automated driving. At present, due to the strict requirements of computing and storage capacity, various GenAI services such as GPT-4 and DeepSeek are deployed on cloud servers, which inevitably results in the high response delay, which is not friendly for providing real-time GenAI services. Mobile-edge-cloud (MEC) collaborative computing would become a competitive technique to empower real-time GenAI services. However, how to utilize these distributed computing resources over MEC networks to meet the requirements of real-time GenAI services remains a challenging issue.

\subsection{Semantic Extraction for Real-Time GenAI Services}
The prompts in real-time GenAI services are basically obtained by sampling the time-varying information-updating process and have different modalities, such as text, image, and voice, which typically follow different mathematical properties. Some information updating processes vary rapidly, while some processes vary slowly. It is noted that there is no need to input the prompts with the same semantics to GenAI models, which would result in redundant computations. For example, in news generation systems, multiple information sources may provide similar events or information. Without semantic deduplication, the system might generate multiple news articles about the same event, resulting in redundant content and wasted resources. On the contrary, identifying and removing semantically similar inputs can ensure that the generated content is more diverse, concise, and timely. As a result, as shown in Fig.~\ref{fig:system_diagram}(b), taking semantic deduplication into account before transmission and computing with GenAI models is essential for real-time information generation.

\subsection{Security Guarantees for Real-Time GenAI Services}
As shown in Fig.~\ref{fig:system_diagram}(c), security guarantees for real-time GenAI services, such as preventing leakage of generative information, is another important issue. An alternative way to prevent information leakage is to deploy the entire GenAI model on local devices. However, since the resources of local devices, whether in terms of computation, memory, or energy, are limited, achieving real-time GenAI may be unrealistic. Therefore, the powerful computing capabilities of edge or cloud servers are essential to support real-time GenAI services. Subsequently, the raw data or intermediate activations on local devices will be transmitted to the edge or cloud server in the process of accessing real-time GenAI services, which raises several security issues. First, since the transmission of raw data or intermediate activations is often over wireless channels, it could be eavesdropped by malicious attackers, resulting in information leakage. Here, the information leakage process should be evaluated and controlled. The second potential risk is the trustworthiness of edge or cloud servers. As GenAI services are deployed over MEC networks, the sophistication of attacks such as man-in-the-middle (MITM) attacks, where an attacker intercepts the communication between the user and the legitimate service provider, increases. A new security vision called zero trust~\cite{zerotrust2010}, which suggests never trust and always verify, could play an important role in preventing such attacks and ensuring the trustworthiness of the provider's identity. In summary, these concerns highlight the need for a comprehensive security approach that not only focuses on real-time GenAI performance, but also prioritizes generative information protection and robust identity verification of service providers to mitigate the potential security issues.

\section{Evolution of Information Aging Metrics Towards Real-Time GenAI}
\label{sec:evolution_aoi}
\begin{table*}
    \centering
    \footnotesize
    \caption{Information-Aging Metrics Towards Real-Time GenAI}\label{tab:aoi_metric}
    \begin{threeparttable}
    \begin{tabular}{|c|c|}
    \hline
    \textbf{Metric} &\textbf{Definition} \\ \hline\hline
    Age of Generative Information (AoGI) & \makecell[l]{the amount of time elapsed since the original information associated with the last generated information \\was sampled}  \\ \hline
    \makecell{Age of Incorrect Information (AoII)\\ in real-time GenAI} & \makecell[l]{the amount of time elapsed since the semantics of original information was changed and the associated\\ generative information was not created yet}  \\ \hline
    \makecell{Age of Leaking Information (AoLI)\\ in real-time GenAI} & \makecell[l]{the amount of time elapsed since the original information associated with the last generative information\\ obtained by the eavesdropper was sampled} \\ \hline
    \makecell{Age of Trust (AoT)} & \makecell[l]{the amount of time elapsed since the last verification of the target user's trust plus the initial age, which\\ depends on the trust level evaluated at that verification} \\ \hline
    \end{tabular}
    \end{threeparttable}
\end{table*}

As mentioned earlier, the information generation and updating processes in real-time GenAI are more diverse than the traditional state updating processes in sensor networks, and thus existing information-aging metrics need to be evolved to match their characteristics. To this end, we discuss several new information-aging metrics in this section.

\begin{figure}
  \centering
  \includegraphics[scale=0.6]{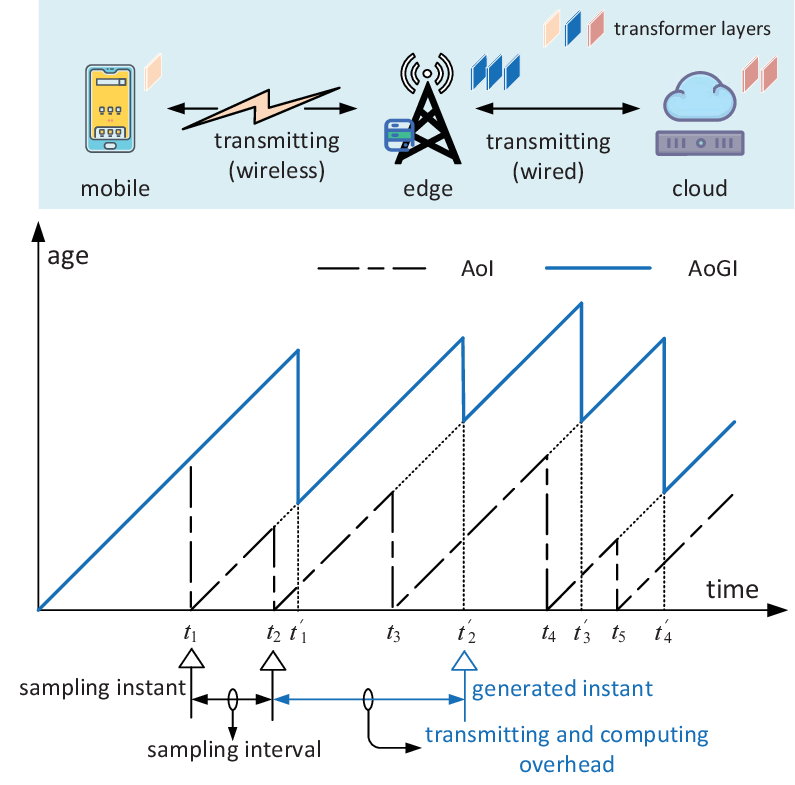}
  \caption{Illustration of age of generative information for real-time GenAI.}\label{fig:aogi}
\end{figure}
\subsection{Age of Generative Information}
As shown in the top part of Fig.~\ref{fig:aogi}, GenAI models are partitioned and then placed in a fragmented fashion across MEC networks to take advantage of the powerful computing and storage capabilities of edge and cloud servers. Consider a real-time GenAI service process in which a time-varying source, such as an ongoing live stream, is continuously sampled. The sampled original information, i.e., the prompt, is then fed into a GenAI model that is partially deployed on a local device for mobile computing. The intermediate activations are then transmitted over a wireless channel to an edge server. Upon receiving these activations, the edge server continues processing using its portion of the GenAI model. As additional model components are deployed on a cloud server, the edge server's outputs are further transmitted to the cloud, where the remaining computations are completed to obtain the generative information. This process is repeated continuously, resulting in an information generation and update process. A natural question is how to evaluate the freshness of the generated information. Ideally, the generative information is expected to be obtained as soon as the corresponding original information is sampled. Age of Information (AoI) is considered as a powerful metric to evaluate the freshness of information. However, existing work on AoI mainly focuses on how to make the target synchronize the state at the source as soon as possible, which is affected by the communication overhead from the source to the target and does not introduce information transformation by computation. In real-time GenAI, the sampled original information, i.e., the prompt, is not only transmitted but also transformed by computation, introducing both communication and computational overhead. In most situations, the computational overhead takes a leading position. Beyond the concept of AoI, we propose the concept of Age of Generative Information (AoGI), which is defined as the elapsed time since the original information associated with the last generated information was sampled. As shown in Fig.~\ref{fig:aogi}, the $i$th original information is sampled at time $t_i$, and the corresponding generative information is generated at time $t'_i$. The age of the original information increases linearly with time and is updated to zero when the new original information is sampled. The AoGI also increases with time, but is updated to the age of the next original information once the corresponding generative information is obtained. In other words, the age of the original information can be considered as the ideal age for the generative information, which means that the generative information can be obtained immediately after its original information has been sampled. To summarize, AoGI takes into account sampling, transmission, and computation. We can use it as a goal to guide the sampling scheme, transmission scheme, and deployment strategy of GenAI models to keep the generative information as fresh as possible.

\subsection{Age of Incorrect Information in Real-Time GenAI}
In real-time GenAI, the original information sometimes does not change often. Even if it changes, the semantics of it may remain unchanged. Motivated by this observation, the semantic extraction for the sampled original information, as shown in Fig.~\ref{fig:system_diagram}(b), is added to prevent the duplicate generation. In this situation, the age of incorrect information (AoII) can be introduced to evaluate the freshness of the generated information~\cite{Maatouk2023The}. As AoI has evolved into AoGI in the context, the definition of AoII can be adjusted as the amount of time that has elapsed since the semantics of the original information was changed and the associated generative information has not yet been created. The adjusted AoII is more suitable than AoGI for communication, computation, and model splitting optimization when introducing semantic extraction into real-time GenAI.

\subsection{Age of Leaked Information in Real-Time GenAI}
When the GenAI models are deployed over MEC networks, we need to transmit the raw data or intermediate activations from the mobile device to the edge server over wireless channels, and there may be some potential malicious eavesdroppers listening to the channels to obtain this information. Therefore, it is necessary to keep the leaked raw data or intermediate activations as old as possible for the eavesdropper. To this end, age of leaked information (AoLI)~\cite{Zheng2024Analysis} can be used to evaluate the staleness of leaked information. It is recommended that the definition of AoLI in real-time GenAI can be adjusted as the time elapsed since the original information associated with the last generative information obtained by the eavesdropper was sampled. When considering generative information leakage issues, AoLI is expected to be maximized while AoGI is minimized, so a fundamental balance between them can be investigated in the future.

\begin{figure}
  \centering
  \includegraphics[scale=0.6]{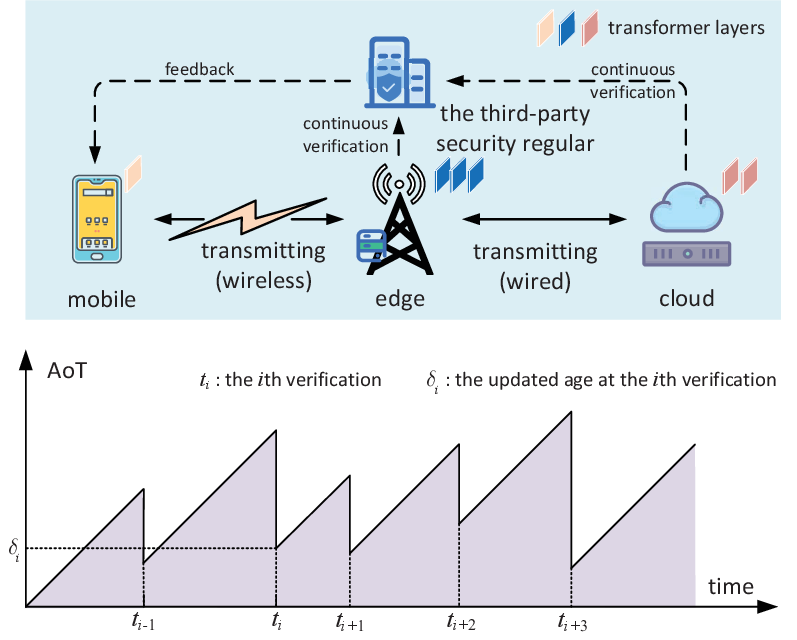}
  \caption{Illustration of the age of trust for continuous identity verification in real-time GenAI.}\label{fig:aot}
\end{figure}
\subsection{Age of Trust}
In addition to preventing information leakage during the data transmission process, the identities of edge or cloud servers providing generative AI services should always be trustworthy. Zero Trust is a novel security assurance framework for device identification verification for future networks, which follows the philosophy of never trust and always verify~\cite{zerotrust2010}. Inspired by this principle, we introduce the concept of age of trust (AoT), which is defined as the elapsed time since the last verification of the target user's trust plus the initial age. The initial age depends on the trust level evaluated at that verification. The higher the trust level, the lower the updated age. An illustration of AoT is shown in Fig.~\ref{fig:aot}, where the $i$th verification is performed at time $t_i$ and the updated age is $\delta_i$. The verification process can be delegated to the third-party security regulars. It can also be performed by the local devices if it is easy and inexpensive. Since essential resources such as computation, communication, and energy resources are required for verification, AoT can be used as a target to find the optimal verification scheme when the resources are constrained. Moreover, the constrained resources could be used for other purposes, such as data transmission and real-time GenAI-based inference. There would be the fundamental trade-off between transmission efficiency, computational efficiency, and trust level.

\section{Optimization of Information Aging Metrics Towards Real-Time GenAI}
\label{sec:opt_aoi}
We have discussed the evolution of information-aging metrics towards real-time GenAI. Due to the dynamic wireless environments as well as the time-varying available computational resources in MEC networks, the corresponding information-aging processes would be random when we deploy the real-time GenAI services over them. It is urgent to investigate how various GenAI techniques, such as model quantization and pruning, and communication techniques should be employed to optimize these information aging processes to meet the diverse requirements of real-time GenAI services.

\begin{figure*}
  \centering
  \includegraphics[scale=0.6]{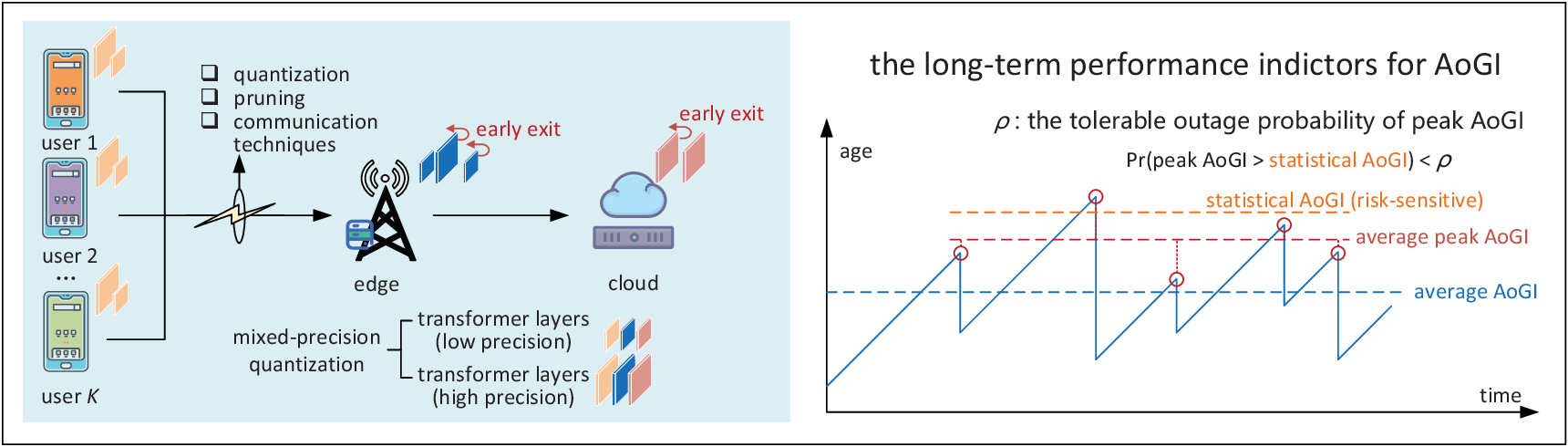}
  \caption{Various techniques for reducing average peak AoGI as well as statistical AoGI towards real-time GenAI services.}\label{fig:long_term_age}
\end{figure*}
\subsection{Average Age or Average Peak Age in Real-time GenAI}
The optimization of information aging metrics, such as AoGI and AoT, towards real-time GenAI will yield various meaningful designs. In the following, we focus on the optimization of AoGI as a representative role. To evaluate the long-term freshness of generative information, the statistics of the AoGI process should be introduced. The average AoGI is considered to be a good performance indicator to evaluate the freshness of generative information from a long-term perspective. However, similar to the average AoI, the derivation of the average AoGI is complicated. Alternatively, it is found that peak AoI can also capture the key characteristics of aging processes~\cite{Yates2021Age}. Therefore, average peak AoGI can be considered as a surrogate for average AoGI to assess the long-term freshness of generative information.

To minimize the average peak AoGI, the allocation of communication and computational resources should be carefully arranged. Regarding the communication resources, since the wireless environments are highly time-varying, it is worth exploring how to allocate the limited communication resources such as transmission power and bandwidth to transmit raw data or intermediate activations as timely as possible. As shown in Fig.~\ref{fig:long_term_age}, we can use quantization and pruning techniques to compress the intermediate activations to reduce the amount of transmitted data, thereby reducing the transmission delay. As for the computational resources, since there are always multiple other computationally intensive services, such as multiple real-time GenAI services or the coexistence with other types of services, running on the edge and cloud servers, it requires a highly efficient allocation scheme in terms of computational resources to reduce the computation delay for real-time GenAI services while satisfying the requirements of other services. In addition, we can use the quantization, model pruning, and knowledge distillation techniques to reduce the model size and thus reduce the dependence of GenAI models on computational resources. It is worth noting that the mixed-precision quantization technique can be adopted to accommodate the different computational capabilities of heterogeneous devices over MEC networks. The high-precision transformer layers are chosen when computational resources are redundant. Otherwise, the low-precision transformer layers can be chosen to reduce the computation delay.

On the other hand, the early exit technique, where we can set multiple early exit points in GenAI models to generate the desired information in advance once the AoGI exceeds a predefined threshold, would be more flexible to control the AoGI. The joint optimization for the allocation of communication and computation resources could have the potential to further reduce the AoGI.An interesting topic is the search for the optimal model-cutting layer when deploying GenAI models over MEC networks. It is expected to find the trade-off between the communication overhead and the computation overhead in the time domain to achieve the minimum average peak AoGI. In addition to efforts to adapt GenAI models, the use of caching techniques to store popular generative information is also a competitive candidate to reduce the average peak AoGI for real-time GenAI services.

In real-world scenarios, multiple real-time GenAI services with different quality-of-service (QoS) requirements may occur simultaneously in a local edge server, where the services with stringent QoS requirements could be given high priority in resource allocation. In this case, the weighted sum of the average AoGI can serve as the objective, and the research of the communication and computation resource allocation scheme to minimize it could be more complicated. It is worth noting that AoGI follows the Markov characteristics, and thus the deep reinforcement learning method can be used as a competitive candidate to search for the optimal scheme.

\begin{figure}
  \centering
  \includegraphics[scale=0.6]{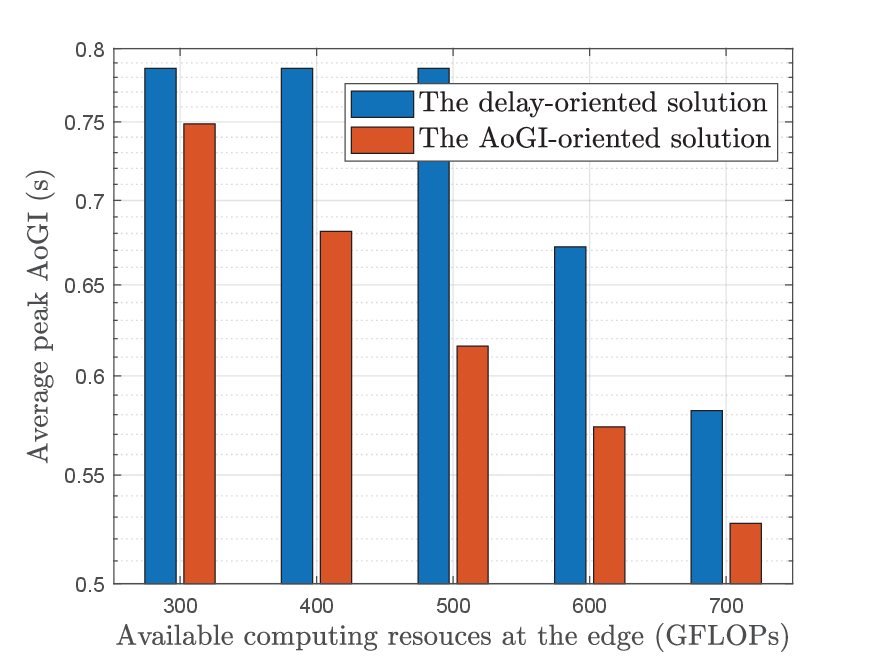}
  \caption{The performance comparison between the delay-oriented solution and the AoGI-oriented solution for model partition.}\label{fig:performance}
\end{figure}

To demonstrate the advantages of the AoGI-oriented solution over the delay-oriented solution, as shown in Fig.~\ref{fig:performance}, we present a simulation where a real-time GenAI service is supported by the mobile-edge collaborative computing integrated with model partitioning techniques. The computational capability of the mobile device, such as a high-performance laptop, is set to 500 GFLOPs. Since many other computational tasks are performed simultaneously in the edge server, the available computational resources for this real-time GenAI service at the edge are set to 300 GFLOPs to 700 GFLOPs. From the figure, it can be seen that the AoGI-oriented model partitioning solution can further reduce the average peak AoGI compared to the delay-oriented model partitioning solution, so that the GenAI model is preferably deployed on the one with high computational capability for a low inference delay. In particular, when the available computational capability of the edge is equal to that of the mobile device, the average peak AoGI is reduced by 25\%.

\subsection{Statistical Age in Real-time GenAI}
Some real-time GenAI applications are risk-sensitive, such as automated driving, where the generative information should always be as fresh as possible. Otherwise, even a small amount of outdated generative information could lead to unwanted consequences. Unfortunately, the average (peak) AoGI is risk-insensitive and cannot reflect the percentage of AoGIs that are too large. To address this issue, the metrics should be introduced in terms of AoGI failure. As mentioned above, peak AoGI can capture the core characteristics of AoGI processes, and we focus on the outage of peak AoGI in the following. First, the minimum achievable peak AoGI for a given outage probability is a risk-sensitive metric to evaluate the freshness of generative information in the long run. However, this metric cannot capture the distribution characteristics of the peak AoGI over the minimum achievable peak AoGI, which may remain around a large value. To fill this gap, the average of the peak AoGI over the minimum achievable peak AoGI comes into view. As its name suggests, it takes into account the average statistic of the crossed peak AoGI. Notably, this metric is equivalent to the metric called conditional value at risk (CVaR) for risk investment in the financial field. The corresponding research has shown that CVaR cannot be computed efficiently~\cite{ahmadi2012entropic}, and proposes to use entropic value-at-risk~(EVaR), which is derived from the Chernoff-Cram\'{e}r bound, to replace it. EVaR is shown to be a tight bound on CVaR. Taking advantage of the Chernoff-Cram\'{e}r bound, EVaR is organized with a form of the moment generating function~(MGF) of the focused variable, and thus we can compute it efficiently. Following the experience in this field, we introduce EVaR to study the failure of AoGI, and for convenience, we call the EVaR in terms of peak AoGI as statistical AoGI. In our previous work, a similar concept called statistical AoGI has been applied to the optimization of state updating in sensor networks~\cite{Xiao2024Statistical}. Statistical AoGI is more suitable than the aforementioned long-term metrics in terms of AoGI for risk-sensitive GenAI applications, especially in guiding the solution design for them. It is worth noting that when the given failure probability is one, the statistical AoGI reduces to the average peak AoGI. In real-world applications, we can adjust the failure probability to meet the different requirements of different risk-sensitive GenAI applications.

\section{Conclusions and Open Challenges}
\label{sec:conclusion}
We investigated the evolution and optimization of information aging metrics for the real-time GenAI services, which focus on the continuous information generation and updating processes. Specifically, considering the diversity and generativity characteristics of the information in real-time GenAI, we introduced two new metrics called Age of Generative Information (AoGI) and Age of Trust (AoT), which are particularly relevant for the empowerment of real-time GenAI services via MEC collaborative computing. To highlight their importance, we discussed the optimizations of these metrics with an emphasis on AoGI, showing the potential of various techniques such as model quantization, model pruning, early exit, etc. to reduce AoGI from a long-term perspective. In particular, with respect to risk-sensitive real-time GenAI services, we introduced a concept, namely statistical AoGI, that can accommodate different sensitivities in risk-sensitive GenAI applications. Research on enabling real-time GenAI services, especially for continuous information generation, is still at an early stage, and there are numerous research opportunities worth exploring. Several open challenges can be further explored, such as:
\begin{itemize}
  \item \textbf{Green Real-Time GenAI Services}: Real-time GenAI services require significant computational resources, resulting in high energy costs and environmental impact. To address this, there is a critical need for green deployment strategies that maintain real-time performance while minimizing carbon emissions. However, fundamental research to clarify the tradeoffs between timeliness, accuracy, and energy efficiency remains unknown.
  \item \textbf{Personalized Real-Time GenAI Services}: Personalized GenAI services should dynamically adapt to individual user preferences, behaviors, and contexts. Guaranteeing timeliness while ensuring deep personalization requires significant computational resources for model updates. It is valuable to discuss how to find a trade-off between model updates and real-time inference. Since personalization relies on sensitive user data, robust privacy protection mechanisms for real-time GenAI services also need to be explored.
  \item \textbf{Collaborative Real-Time GenAI Services}: It can be viewed as a vertical perspective on enabling real-time GenAI services via MEC collaborative computing. When there are multiple fragments of GenAI models distributed across different mobile devices, various horizontal collaboration techniques, such as device-to-device communication, would have the potential to facilitate the real-time GenAI services. There is a critical need to discuss the corresponding performance gains and costs to identify when the horizonal collaboration techniques can serve as a cost-effective way to enhance the real-time GenAI services.
\end{itemize}

\bibliographystyle{IEEEtran}
\bibliography{References}

\end{document}